# Behavior of Electric Current Subjected to ELF Electromagnetic Radiation

Fran De Aquino

Physics Department,
Maranhao State University,
S.Luis/MA, Brazil.

Gravitational effects produced by ELF electromagnetic radiation upon the electric current in a conductor are studied. An apparatus has been constructed to test the behavior of current subjected to ELF radiation. The experimental results are in agreement with theoretical predictions and show that ELF radiation can cause transitory interruptions in electric current conduction.

## 1. INTRODUCTION

In a recent paper[1] we have shown that the *gravitational mass*, $m_g$, and the *inertial mass*, $m_i$, are correlated by a dimensionless factor, which depends on the incident(or emitted) radiation upon the particle. The general expression of correlation can be written as:

$$m_g = m_i - 2\left\{\sqrt{1+\left\{\frac{aD}{m_i c}\sqrt{\frac{\mu\sigma}{4\pi f^3}}\right\}^2} - 1\right\}m_i \quad [1]$$

where $D$ is the *power density* of the incident( or emitted) radiation; $f$ is the frequency of the radiation; $a$ is the area of the surface of the particle of mass $m_i$; $\mu$ and $\sigma$ are respectively, the *permeability* and the *conductivity* of the medium around the particle, in which the incident radiation is propagating. For an *atom* inside a body, the incident(or emitted) radiation upon the atom will be propagating inside the body, and consequently, $\sigma = \sigma_{body}$, $\mu = \mu_{body}$.

Equation[1] shows that, elementary particles (mainly *electrons*) can have their *gravitational masses* strongly reduced by means of Extremely-Low Frequency (ELF) radiation.

In this paper, we will study the gravitational effects of ELF radiation upon the electric current through a conductor.

## 2. THEORY

Let us consider an electric current $I$ through a conductor submitted to ELF radiation with power density $D$ and frequency $f$.

Under these circumstances the *gravitational mass* $m_{ge}$ of the *free electrons* (electric current), according to Eq.[1], is given by

$$m_{ge} = m_e - 2\left\{\sqrt{1+\left\{\frac{a_e D}{m_e c}\sqrt{\frac{\mu\sigma}{4\pi f^3}}\right\}^2} - 1\right\}m_e \quad [2]$$

Here, $\mu$ and $\sigma$ are respectively, the *permeability* and the *conductivity* of the *conductor*; $m_e = 9.11\times10^{-31} kg$.

According to Eq.[2], $m_{ge} = 0$ when

$$D = \frac{m_e c}{a_e}\sqrt{\frac{5\pi f^3}{\mu\sigma}} \quad [3]$$

Consequently, if

$$D \gg \frac{m_e c}{a_e} \sqrt{\frac{5\pi f^3}{\mu \sigma}} \quad [4]$$

*gravitational* mass of the free-electrons becomes *strongly negative*. For this circumstance, *the gravitational attraction* between the electrons, $F = -G m_{ge} m'_{ge} / r^2$, will *increase* strongly. As a result, the *free-electrons gas* is compressed, and consequently, the cross-section, $S_I$, which is crossed by the current, *decreases*. When the area $S_I$ decreases, the resistance to the current $R = l/\sigma S_I$, *increases*. Therefore, a strong attraction between the free-electrons increases strongly the resistance $R$ and can *interrupt* the current.

Assuming that the *radius* of the electron[2] (region which electron is "concentrated") is

$$r_e = (1/4\pi \varepsilon_0)(e^2/m_e c^2) = 2.8 \times 10^{-15} m,$$

$a_e = 4\pi r_e^2 = 9.8 \times 10^{-29} m^2$ and $\mu \cong \mu_0; \sigma = 5.8 \times 10^7 S/m$ (Copper conductors), then the condition[4] becomes

$$D \gg 10^6 f^{\frac{3}{2}} \quad [5]$$

This means that ELF radiation fluxes of frequency $f = 10^{-6} Hz$, for example, with power density sufficiently strong ($D \gg 10^{-3} W/m^2$) can cause *current interruptions* in electric circuits.

It is important to note that current interruptions in *ferromagnetic conductors* of high permeability request a smaller value of $D$ ($\ll 10^{-3}$ W/m$^2$). See Eq.[2]; if $\mu$ increases it is necessary a smaller $D$.

The ELF radiation must to reach most of the free-electrons. We know that there is *one* free-electron by cooper atom. On the other hand, the number of atoms/m$^3$ of copper is

$$n = \frac{N_0 \rho}{A} =$$
$$= \frac{(6.02 \times 10^{26} atoms/kmole)(8900 kg/m^3)}{63.54 kg/kmole} =$$
$$= 8.4 \times 10^{28} atoms/m^3$$

Thus, there are $8.4 \times 10^{28}$ free-electrons/m$^3$. Therefore, it is necessary $10^{29}$ photons/m$^2$ to reach *all* the free-electrons inside 1m$^3$ of copper. However,

$$10^{29} photons/m^2 \equiv \frac{1 photon}{10^{-29} m^2}$$

Consequently, the power density $D$ of the ELF radiation must be:

$$D = \frac{hf^2}{10^{-29} m^2} \approx 10^{-5} f^2 \quad [6]$$

Obviously, condition[5] overcomes the condition[6].

From Electrodynamics, we know that a radiation of frequency $f$ propagating within a material with electromagnetic characteristics $\varepsilon$, $\mu$ and $\sigma$ has the amplitudes of its waves attenuated by $e^{-1} = 0.37$ (37%) when it penetrates a distance $z$, given by [3]

$$z = \frac{1}{\omega \sqrt{\frac{1}{2} \varepsilon \mu \left( \sqrt{1 + (\sigma/\omega\varepsilon)^2} - 1 \right)}} \quad [7]$$

From the equation above we conclude that there is minimal shielding for radiation of frequency $f = 10^{-6} Hz$.

## 3. EXPERIMENTAL

Consider the circuit in Fig.1 where series connection of an ammeter and a 300W-220V lamp are connected at an AC source by means of a *annealed iron wire* ( 99.98% Fe; $\mu = 350,000 \mu_0; \sigma = 1.03 \times 10^7 S/m$). We have opted by *ferromagnetic conductor* of *highest permeability* because, as we have seen, a smaller value of $D$ (and consequently, a smaller current through the antenna ) is requested.

A *segment* of the annealed iron wire is subjected to ELF radiation of frequency $f = 10^{-6} Hz$ radiated from a *spiral antenna* as shown in Fig.1. The spiral antenna is a *half-wave dipole*, *encapsulated by an iron disc* ( 99.95% Fe; $\mu_i = 5,000\mu_0$; $\sigma_i = 1.03 \times 10^7 S/m$ ).

The iron surrounding the dipole increases its inductance $L$. However, for series RLC circuit the *resonance frequency* is $f_r = 1/2\pi\sqrt{LC}$, then when $f = f_r$,

$$X_L - X_C = 2\pi f_r L - \frac{1}{2\pi f_r C} = \sqrt{\frac{L}{C}} - \sqrt{\frac{L}{C}} = 0.$$

Consequently, the impedance of the antenna, $Z_{ant}$, becomes *purely resistive*, i.e.,

$$Z_{ant} = \sqrt{R_{ant}^2 + (X_L - X_C)^2} = R_{ant} = R_r + R_{ohmic}.$$

For $f = f_r = 10^{-6} Hz$ the length of the dipole is

$$\Delta z = \lambda/2 = (v/f)/2 = \sqrt{\pi/\mu_i \sigma_i f} = 6.97m$$

here $v = c/n_r$ is the speed of ELF radiation in the medium; $n_r$ is the index of refraction.

The table 1 presents the experimental results obtained for current $I$ (current through the annealed iron wire) as a function of the current $I_a$ (current through the antenna; $1$-$3A$ ). The current $I$ is *interrupted* when $I_a \sim 2.48A$.

## 4.CONCLUSION

With the new technology of transmitter antennas encapsulated with ferromagnetic materials[4], applied to Radar systems, it is possible to generate pulses of ELF radiation with a frequency $f = 10^{-6} Hz$ and power density $\sim 0.1 \ W/m^2$ at distances ~100km from the transmitter ( therefore we have $D \gg 10^{-3} W/m^2$ ). In addition, one can build a device similar to a *laser* using ELF radiation instead of visible light waves.

These fluxes can be very dangerous to satellites, aircrafts, missiles, transmission lines, power plants and also perhaps to *Humans* causing transitory interruptions in the transmission of the *nervous system impulses* ( ionic current ). However there are also benefits; ELF radiation can be very useful as "*electronic anesthesia*".

## REFERENCES

1. De Aquino, F.(2000)"*Gravitation and Electromagnetism: Correlation and Grand Unification*", Journal of New Energy , vol.**5**, no2 , pp.76-84. Los Alamos National Laboratory preprint no.gr-qc/9910036.
2. Alonso, M., Finn, E.J.(1972) *Física*, Ed. Edgard Blücher, p.149. Translation of the edition published by Addison-Wesley (1967).
3. Quevedo,C.P.(1978)*Eletromagnetismo* McGraw-Hill,p.269-270.
4. De Aquino, F.(2000) "*Possibility of Control of the Gravitational Mass by means of Extra-Low Frequencies Radiation*", Los Alamos National Laboratory preprint no.gr-qc/0005107.
   De Aquino, F.(2002) " *Correlation between Gravitational and Inertial Mass: Theory and Experimental Test* ", Los Alamos National Laboratory preprint no.physics/0205089.

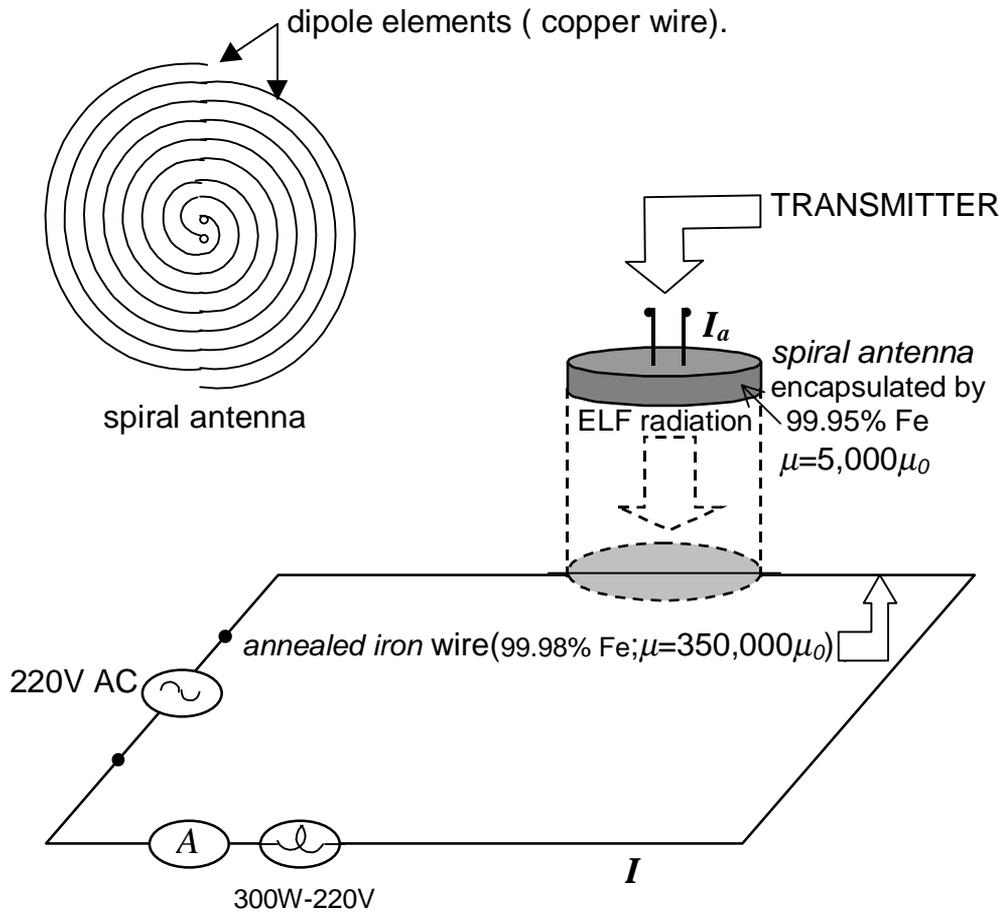

Fig.1- Schematic diagram of the experimental set-up

| $I_a$ experim. ( A ) | $D$ theory ( W/m$^2$ ) | $m_{ge}/m_e$ theory | $I$ experim. ( A ) |
|---|---|---|---|
| 0 | 0 | 1 | 1.358 |
| 0.5 | 3.1 E-06 | 0.59 | 1.358 |
| 1.0 | 1.2 E-05 | -2.68 | 0.191 |
| 1.5 | 2.8 E-05 | -9.13 | 0.018 |
| 2.0 | 4.9 E-05 | -18.34 | 0.004 |
| 2.5 | 7.7 E-05 | -30.30 | 0.000 |
| 3.0 | 6.9 E-04 | -44.91 | 0.000 |

Table 1